\documentclass[twocolumn,aps,prb,superscriptaddress,a4paper,showpacs,showkeys]{revtex4-1}

\setcitestyle{numbers,square}

\usepackage{graphicx}
\usepackage{bm,amsmath,amssymb,dcolumn}
\usepackage{url,cancel}
\usepackage{type1cm}
\usepackage[dvips]{color}
\usepackage[normalem]{ulem}

\newcommand{\mm}[1]{\mbox{$#1$}}

\newcommand{\ke}{$K$~edge}
\newcommand{\kkr}{{\sc sprkkr}}
\newcommand{\wtk}{{\sc wien2k}}

\newcommand{\ea}{{\it et al.}}

\begin{document}

\title{Ca and S $K$ edge XANES of CaS calculated by different methods:
  influence of full potential, core hole and Eu doping }

\author{Ond\v{r}ej \v{S}ipr}
\email{sipr@fzu.cz}
\homepage{http://www.fzu.cz/~sipr} 
\affiliation{Institute of Physics, Czech Academy of Sciences,
  Cukrovarnick\'{a}~10, CZ-162~53~Prague, Czech Republic }
\affiliation{University of West Bohemia, Univerzitn\'{\i} 8,
  CZ-306~14~Pilsen, Czech Republic}

\author{Wilayat Khan}
\affiliation{University of West Bohemia, Univerzitn\'{\i} 8,
  CZ-306~14~Pilsen, Czech Republic}
\affiliation{Department of Physics, University of Lahore, Gujrat
  Campus, Gujrat, Pakistan} 

\author{Yves Joly}
\affiliation{Univ.\ Grenoble Alpes, CNRS, Grenoble INP, Institut
  N\'{e}el, 38000 Grenoble, France}

\author{J\'{a}n Min\'{a}r}
\affiliation{University of West Bohemia, Univerzitn\'{\i} 8,
  CZ-306~14~Pilsen, Czech Republic}


\date{\today}


\begin{abstract}
Ca and S \ke\ spectra of CaS are calculated by the full-potential
Green's function multiple-scattering method, by the FLAPW method, and
by the finite difference method.  All three techniques lead to similar
spectra.  Some differences remain close to the edge, both when
comparing different calculations with each other and when comparing
the calculations with earlier experimental data.  We find that using
the full potential does not lead to significant improvement over the
atomic spheres approximation and that the effect of the core hole can
be limited to the photoabsorbing atom alone.  Doping CaS with Eu will
not affect the Ca and S \ke\ XANES of CaS significantly but may give
rise to a pre-edge structure not present for clean CaS.
\end{abstract}




\maketitle  


\section{Introduction}

\label{sec-intro}

Cubic calcium sulfide (CaS) is an important test material for studying
the physical mechanism behind the processes in light emitting diods
(LED's).  This is especially true for CaS doped with rare earth metals
such as Eu \cite{TS+93,ZLZ+05,JW+07,Hua+15}.  A detailed understanding
of electronic and spectroscopic properties of both clean and doped CaS
is desirable. This applies also to the x-ray absorption spectra.

The accuracy of x-ray absorption near edge structure (XANES)
calculations has improved a lot during recent years.  Among various
factors affecting the accuracy of the calculation are the assumed
shape of the potential (the full potential {\it vers.}\ the muffin-tin
potential) and the treatment of the core hole.  One of the goals is to
achieve a balance between the accuracy and computational demands.
Therefore it is important to disentangle the physical effects from
effects related to particular methods or codes.

The S \ke\ XANES of CaS was investigated experimentally and
theoretically in the past \cite{FFS+02,KSS+04,APG+09}.
High-resolution experimental data for both the Ca and S $K$~edge XANES
were published in a recent study of Xu \ea\ \cite{XLC+13}.  They
performed also calculations, either by the Green's function
multiple-scattering method as implemented in the {\sc fdmnes} code
\cite{fdmnes-code,BJ+09} and in the {\sc feff} code
\cite{feff-code,RKV+10}, or by the finite-difference method (FDM) as
implemented in the {\sc fdmnes} code.  The multiple-scattering
calculations were done relying on the muffin-tin approximation while
the finite-difference method calculations were done using a full
potential (no shape approximation).  Xu \ea\ \cite{XLC+13} obtained
quite a good agreement between theory and experiment close to the
edge.  For the Ca \ke, they found that the full-potential calculation
led to a significantly better agreement between theory and experiment
than the muffin-tin calculation.  This can be considered as a surprise
because it has been believed (based on previous experience) that
full-potential effects are not very important for highly symmetric
cubic structures \cite{HE+99}.  On the other hand, even the
full-potential {\sc fdmnes} calculations of Xu \ea\ \cite{XLC+13} failed
to reproduced some small but well-defined peaks occurring at 35~eV
above the Ca \ke\ and at 23~eV above the S \ke.  Green's function
multiple-scattering calculations of Kravtsova \ea\ \cite{KSS+04} did not
reproduce the peak at 23~eV above the S \ke\ either.
Xu \ea\ \cite{XLC+13} tentatively attributed this failure to radiation
damage or to an unspecified charge transfer effect.

Given the importance of CaS for research on LED materials, it is
desirable to understand its electronic structure correctly. In
particular it would be interesting to understand more deeply the
surprising findings which follow from the study of Xu \ea\ (need for a
full-potential treatment for the Ca \ke\ together with inability of
both the multiple-scattering and FDM methods to reproduce some peaks
relatively well above the edge).  From a more general point of view,
it is desirable to calculate XANES of CaS by different methods and to
compare the results, so that differences which stem from using
different calculational methods can be separated from differences
which stem from using different physical approximations.

As mentioned in the beginning, the practical interest in CaS (and 
analogous sulfides) stems to a large extent from the fact that when
doped with rare earth atoms these systems form an important class of
LED materials.  The sulfides are particularly interesting because of
their unusual dependence of the thermal quenching of the luminescence
on the dopant \cite{Dor+05}.  To facilitate further experimental
research on the electronic structure of doped sulfides, it would be
instructive to investigate how the Ca and S \ke\ XANES change if CaS
is doped by a rare-earth metal such as Eu.

To achieve these goals, we performed XANES calculations for CaS using
a fully relativistic full potential Green's function method as
implemented in the {\sc sprkkr} code \cite{sprkkr-code,EKM11}, using a
full potential linearized augmented plane wave (FLAPW) method as
implemented in the {\sc wien2k} code \cite{Blaha+01}, and using an FDM
as implemented in the {\sc fdmnes} code \cite{fdmnes-code,BJ+09}.
Apart from comparing the results of different codes and methods, we
investigated the effect of the spherical potential approximation and
of various technical procedures to include the core hole within the
final state approximation.  We found that all three methods lead to
similar spectra which are in a good agreement with experiment --- even
in those energy regions where problems were observed earlier.
Nevertheless, some differences between theoretical and experimental
spectra remain within the first 10~eV above the edge.  We found that
full-potential calculations do not lead to significant improvement
over calculations done within the atomic spheres approximation. The
technical details how the final state approximation is implemented are
not important.  As concerns the effect of Eu doping, we found that it
will not introduce significant changes in the Ca and S \ke\ spectra.


\section{Methods}

\label{sec-methods}

We considered CaS in a B1 or rock salt structure (space group 225,
Pearson symbol cF8), with lattice constant $a_{0}$=5.595~\AA.  The
Green's function or multiple-scattering calculations were done using
the {\sc sprkkr} code.  The potential was either without any
restriction (full potential mode) or subject to the atomic spheres
approximation (ASA).  The ASA is similar to the muffin-tin
approximation but the atomic spheres are overlapping and fill the
whole space so that there is no interstitial region.  Even though
formally less accurate than the muffin-tin approximation, the ASA
often yields better results than the ``pure'' muffin-tin
approximation; it has been suggested that the overlap between the
spheres partially compensates for the defects of the spherical
approximation~\cite{ZA+09}.  To achieve a better filling of the space,
we introduced empty spheres (or Voronoi polyhedra, in the full
potential mode) at the interstitial sites.  The \kkr\ calculations
were done in a fully relativistic mode (solving the Dirac equation),
both in the reciprocal space and in the real space.  For reciprocal
space calculations, the $\bm{k}$-space integration was carried out via
sampling on a regular $\bm{k}$-mesh, using a grid of
45$\times$45$\times$45 points in the full Brillouin zone.  When
calculating the spectra in a real-space mode, a cluster of 123 atoms
(radius of 8.7~\AA) was employed, using the same self-consistent
potential as in the reciprocal-space calculations. The maximum angular
momentum used for the multipole expansion of the Green's function was
$\ell_{\text{max}}$=3.  The full-potential mode relied on the shape
functions, meaning that the full potential and ASA calculations were
performed using the same formalism: in both cases, the same
multiple-scattering equation is solved \cite{Zel87a,HZE+98}.

The \wtk\ calculations are based on the FLAPW method.  The
wave-functions inside the muffin-tin spheres were expanded in
spherical harmonics up to the maximum angular momentum
$\ell_{\text{max}}$=10. The wave functions in the interstitial region
were expanded in plane waves, with the plane wave cutoff chosen so
that $KR_{\text{MT}}$=7 ($R_{\text{MT}}$ represents the smallest
muffin-tin radius and $K$ is the magnitude of the largest wave
vector).  The muffin-tin radii were 2.42~a.u.\ for Ca atoms and
2.31~a.u.\ for S~atoms.  The $\bm{k}$-space integration was performed
via a modified tetrahedron integration scheme, using 1000
$\bm{k}$-points in the full Brillouin zone distributed according to
the 10$\times$10$\times$10 Monkhorst-Pack grid.

The finite difference method solves the Schr\"{o}dinger equation by
discretizing it over a grid of points in the volume where the
calculation is made.  The {\sc fdmnes} code uses in fact a mixed
approach: the wave functions are expanded in spherical waves within
small spheres around the nuclei (similar as in the FLAPW method) while
a standard FDM calculation is done in the interstitial region
\cite{Jol+01}.  The potential used for FDM calculations is
self-consistent as concerns its spherical aspects: it is obtained so
that first a self-consistent calculation using the multiple-scattering
method within the muffin-tin approximation is done and then one last
iteration is performed to obtain the final potential without any
shape restrictions \cite{BJ+09}.  The spectra calculated by the {\sc
  fdmnes} code for this study were obtained for clusters of 81 atoms
(7~\AA~radii).

The calculations were done relying the local density approximation
(LDA) to treat the exchange and correlation effects.  Additionally,
the {\sc fdmnes} calculations (working usually on a larger energy
range) employ a real energy-dependent self-energy
\cite{HL69,LB77}. One can thus expect slight shifts in the peak
positions between the {\sc sprkkr} and {\sc wien2k} calculations on
the one hand and the {\sc fdmnes} calculations on the other.

Calculations for Eu-doped CaS were done using the \wtk\ code (FLAPW
method).  We employed a supercell of 64 atoms, with one Ca atom
substituted by Eu.  The muffin-tin radii were 2.50~a.u.\ for Ca atoms,
2.18~a.u.\ for S~atoms, and 2.50~a.u.\ for the Eu atom.  The structure
was relaxed (keeping the lattice constant intact) to account for the
changes of the bond lengths around Eu.  While the electronic structure
of clean CaS can be described within the LDA, for Eu-doped CaS this
would be inappropriate because of the strongly correlated electrons in
the Eu orbitals.  Therefore we used the LDA+$U$ method as implemented
in the \wtk\ code, with the effective on-site Coulomb interaction
parameter $U$=8.0 eV applied to the Eu 4$f$ electrons. This value
reproduces properly the splitting between the spin up and spin down
electrons for the Eu$^{2+}$ ion.

The raw theoretical spectra were convoluted by a Lorentzian with an
energy-dependent full width at half maximum set as \mm{w(E) =
  w_{\text{core}} + 0.05\times(E-E_{\text{edge}})}.  The constant part
$w_{\text{core}}$ accounts for the finite core hole lifetime [0.77~eV
  for the Ca \ke\ and 0.52~eV for the S \ke\ \cite{CP01}], the
energy-dependent part accounts for the finite lifetime of the ejected
photoelectron \cite{MJW82}.


\subsection{The core hole treatment}

\label{sec-coremethod}

The presence of the core hole can be a major factor in XANES.
However, in case of the \ke\ spectra, where the electron is
ejected into delocalized $p$ states, its influence can be
often described within the final state approximation.  It consists in
evaluating the spectra for electron states which have relaxed to the
presence of the core hole.  Usually, the originally core electron is
put into the valence band: technically this maintains the charge
neutrality, physically this simulates the screening of the core
hole.

For the \kkr\ calculations one can conveniently employ the embedded
impurity formalism: first the electronic structure of the system
without a core hole is calculated and then a Dyson equation is solved
for an embedded impurity cluster centered around the atom with a core
hole \cite{MBS+06}.  The embedded cluster defines the region where the
electronic structure is allowed to relax to the presence of the core
hole; beyond it there is an unperturbed host.  As we have formally an
infinite system with just a single core hole, one does not need to
care about possible interference of the holes.

The embedded cluster around the photoabsorbing atom contains 19 atoms
in this study.  That should be sufficient because the core hole is
usually screened quite efficiently \cite{Zel88}.  Indeed, one can make
an approximation that the core hole does not influence the electronic
structure of neighboring atoms at all, i.e., that its impact is
limited just to the photoabsorbing atom.  Such situation can be
modeled by the single-site impurity formalism: the system is
treated as a substitutional alloy where the concentration of the
normal (ground-state) atoms is 100~\% and the concentration of the
photoabsorbing atoms (with the core hole) is 0~\%. Such a system can
be treated by a computationally efficient coherent potential
approximation (CPA).  As the CPA gives access to type-resolved
quantities, the XANES generated at the atom with the core hole can be
calculated even if its concentration is formally zero.

Physically the single-site approach means that the influence of
neighboring atoms on the atom with the core hole is taken into account
while the influence of the core hole atom on the other atoms is
neglected.  In other words, the core hole is perfectly screened.  The
difference between the embedded-cluster and single-site approaches to
the core hole could be also viewed as a difference between the time
scales in which the electrons respond to the disturbance at the
photoabsorbing atom.  The embedded cluster approach assumes that the
electrons adjust instantaneously to the perturbation.  The single-site
approach assumes that the electrons at neighboring atoms are latent
with respect to the photoabsorption process.

For the \wtk\ calculations, we employed supercells of 64 atoms and
assigned the core hole to one of them.  This means that the electronic
structure of the atoms which are close to the photoabsorbing atom is
affected by the core hole and, at the same time, the big size of the
supercell makes the interaction between two excited atoms practically
negligible.  Such a approach is equivalent to the embedded cluster
approach described above.  On the other hand, for the {\sc fdmnes}
calculations we accounted for the core hole at the photoabsorbing atom
only \cite{BJ+09}.  That is similar to the single-site CPA-like
approach.

Apart from the final state approximation, another commonly used
procedure to deal with the core hole is the Slater transition state
method.  It is similar to the final state approximation, except that
only a half of an electron is transferred from the core level to the
valence band.  The incentive for using this approach is that in this
way one gets a good estimate of transition energies.

CaS is an insulator, so the influence of the core hole should be more
significant than in metals.  To estimate the importance of the core
hole effect and also to check the robustness of different
computational approaches, we performed the calculations using the
ground state potential, the Slater transition state approach and the
final state approximation.


\section{Results: C\MakeLowercase{a} and S  $K$ edge XANES of
  C\MakeLowercase{a}S} 

\label{sec-results}


\subsection{Comparison between theory and experiment}

\label{sec-exper}

\begin{figure}
\caption{Experimental Ca \ke\ (upper panel) and S \ke\ (lower panel)
  XANES of CaS \cite{XLC+13} together with theoretical spectra
  obtained using the full-potential multiple-scattering method
  (\kkr\ code), the FLAPW method (\wtk\ code), and the full-potential
  FDM  ({\sc fdmnes} code).}
  \includegraphics[viewport=0.3cm 0.5cm 9.2cm 14.7cm,width=85mm]
                  {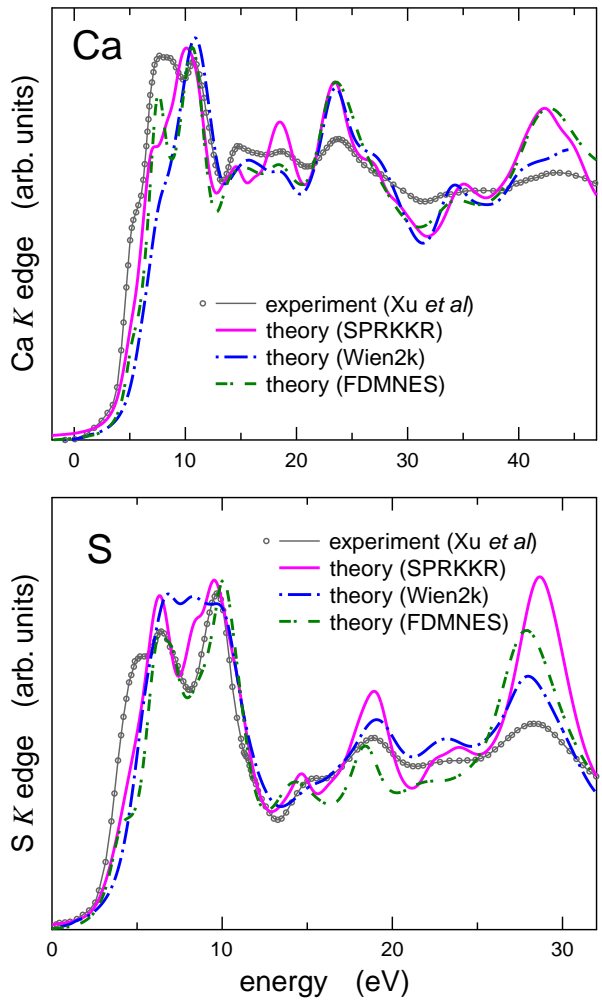}%
\label{fig-exper}
\end{figure}

We begin by comparing calculations with experiment.  This is done in
Fig.~\ref{fig-exper}.  The experiment is taken from the work of
Xu \ea\ \cite{XLC+13}.  We compare it with three different full
potential calculations.  The core hole was included within the final
state approximation, as described in Sec.~\ref{sec-coremethod}.  Note
that for the \kkr\ calculations the final state approximation was
implemented using the embedded impurity cluster method in this
section. 

One can see from Fig.~\ref{fig-exper} that even though all three
calculations use essentially the same physical constraints (full
potential, core hole via the final state approximation), there are
differences in the results.  Some of these differences are not really
significant.  Others are more relevant, especially close to the edge
(($E<10$~eV), where none of the calculations reproduces the experiment
very accurately. Nevertheless, the {\sc fdmnes} calculation is closer
to the experiment in this region than the \kkr\ or \wtk\ calculations.
The reason for this is unclear.


\subsection{Comparing full-potential and ASA calculations}

\label{sec-FPASA}

\begin{figure}
\caption{Theoretical Ca \ke\ (upper panel) and S \ke\ (lower panel)
  XANES of CaS obtained in the full-potential (FP) mode either in the
  real space (for a cluster of 123 atoms) or in the reciprocal space,
  together with XANES obtained for a spherical potential (ASA) in the
  reciprocal space. The calculations were done using the \kkr\ code,
  the core hole was not taken into account.  Experimental XANES of
   Xu \ea\ \cite{XLC+13} is shown for comparison. }
\includegraphics[viewport=0.3cm 0.5cm 9.2cm 11.5cm,width=85mm]
                {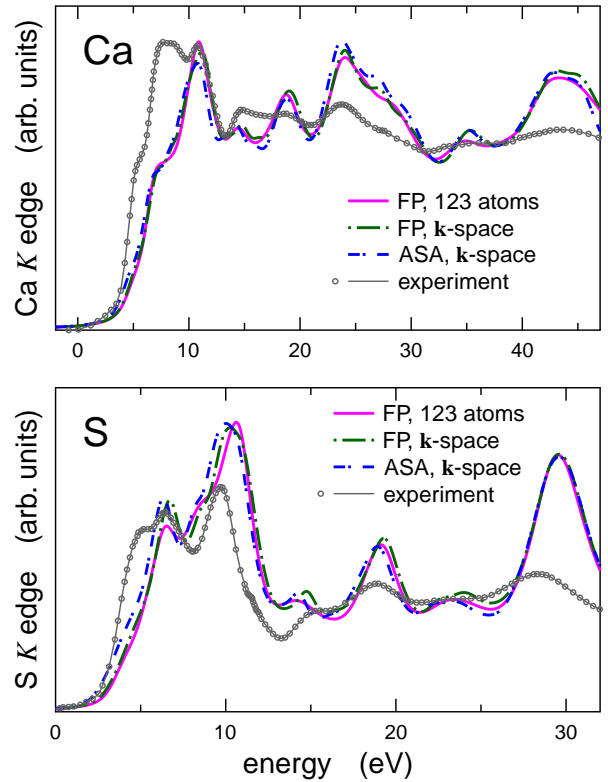}%
\label{fig-FPASA}
\end{figure}

Fig.~\ref{fig-FPASA} compares theoretical Ca and S \ke\ XANES of CaS
obtained for the full potential (FP) and for the spherical potential
(ASA).  The calculations were done within the Green's function
multiple-scattering formalism, by the \kkr\ code.  The core hole was
ignored.  The spectra were calculated in the $\bm{k}$-space.
Additionally, the FP spectra were calculated also in the real-space
(123 atoms in the cluster, 8.7~\AA\ radius).  The experimental spectra
of Xu \ea\ \cite{XLC+13} are shown as well so that one can better
judge the significance of the changes induced by different theoretical
modes. 

One can see that full-potential and ASA calculations lead to very
similar spectra in our case --- even in the near-edge region.  This is
different from the results of Xu \ea\ \cite{XLC+13}: they found that
for the Ca \ke\ spectrum there is a small but significant difference
between their Green's function multiple-scattering calculations done
for a spherical muffin-tin potential and their FDM calculations done
for a full potential.

For both the real-space and the reciprocal-space calculations, some
technical issues have to be dealt with.  For the real-space
calculations, the question is whether the cluster is big enough.  For
the reciprocal-space calculations, the questions are about the
$\bm{k}$-mesh used for the Brillouin zone integration and about the
convergence of the Ewald summation used to evaluate the structure
constants. The real-space and the $\bm{k}$-space spectra shown in
Fig.~\ref{fig-FPASA} are in a good agreement, confirming that in both
cases the technical issues have been handled properly.


\subsection{Dealing with the core hole}

\label{sec-core}

\begin{figure}
\caption{Theoretical Ca \ke\ (upper panel) and S \ke\ (lower panel)
  XANES of CaS obtained when the core hole was accounted for via the
  final state approximation, when the core hole was included via the
  Slater transition state approach, and when it was ignored.  The
  calculations were done using the \kkr\ code in the ASA mode, via the
  embedded cluster formalism.   Experimental XANES of
  Xu \ea\ \cite{XLC+13} is shown for comparison. }
\includegraphics[viewport=0.0cm 0.0cm
  9.0cm 13.5cm,width=85mm] {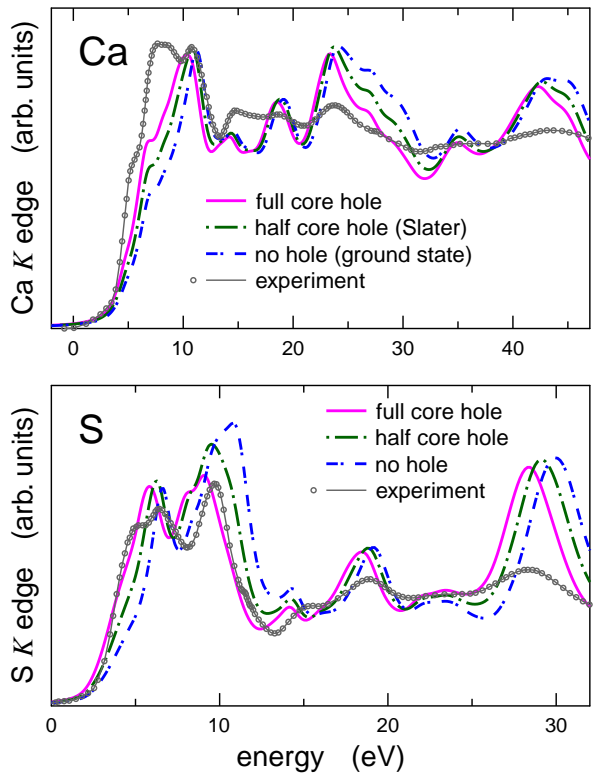}
\label{fig-core}
\end{figure}

Fig.~\ref{fig-core} shows how the spectra change if the core hole is
switched on: one can see the results obtained within the final state
approximation (full core hole), within the Slater transition state
approximation (half core hole) and for the ground state (no core
hole).  The spectra were obtained using the \kkr\ code for a spherical
ASA potential, using the embedded impurity cluster formalism.  One can
see that the core hole does not introduce new features in the spectra
but changes the intensities of peaks and shoulders.  In particular, it
enhances the features close to the absorption edge, in accordance with
common experience \cite{WCA+90,SMS+97,SR+10}.

\begin{figure}
\caption{Theoretical Ca \ke\ (upper panel) and S \ke\ (lower panel)
  XANES of CaS obtained when the core hole was accounted for via the
  final state approximation relying either on the embedded impurity
  cluster formalism or on the single-site impurity approach.
  The calculations were done using the \kkr\ code in the ASA mode.
  Experimental XANES of
   Xu \ea\ \cite{XLC+13} is shown for comparison. }
\includegraphics[viewport=0.0cm 0.0cm 9.0cm 13.5cm,width=85mm]
                {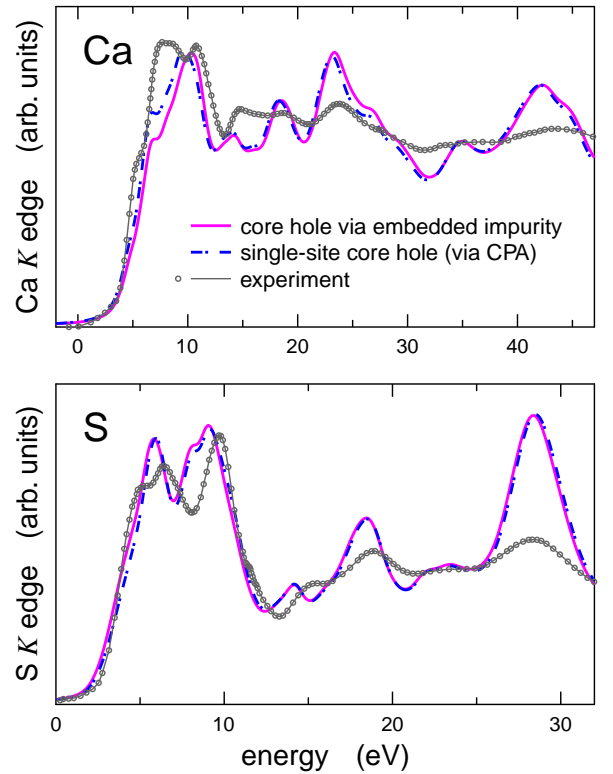}
\label{fig-embed}
\end{figure}

Comparison between spectra calculated using the embedded impurity
cluster method and using the single-site impurity CPA-like approach is
shown in Fig.~\ref{fig-embed}.  Again, the spectra were obtained for
the spherical ASA potential.  Physically, the difference between both
processes is that for the embedded impurity cluster formalism the core
hole affects not only the electronic structure of the photoabsorbing
atom but also of further eighteen atoms in the two nearest
coordination spheres, while for the single-site impurity formalism the
core hole affects just the electronic structure of the photoabsorbing
atom alone.  It is obvious from Fig.~\ref{fig-embed} that spectra
obtained for both approaches lead to very similar spectra, meaning
that the influence of the core hole on the neighboring atoms can be
neglected.


\subsection{Influence of the Eu impurity}

\label{sec-eucas}

\begin{figure}
\caption{Theoretical Ca \ke\ (upper panel) and S \ke\ (lower panel)
  XANES of Eu-doped CaS evaluated for those Ca or S atoms which are
  nearest to the Eu impurity.  The calculations were done by the
  \wtk\ code, either within the LDA framework ($U$=0~eV) or using the
  LDA+$U$ method ($U$=8~eV).  For the Ca \ke, both spectral curves are
  practically identical.  Results for clean CaS are shown for
  comparison. } \includegraphics[viewport=0.0cm 0.0cm 9.0cm
  13.5cm,width=85mm] {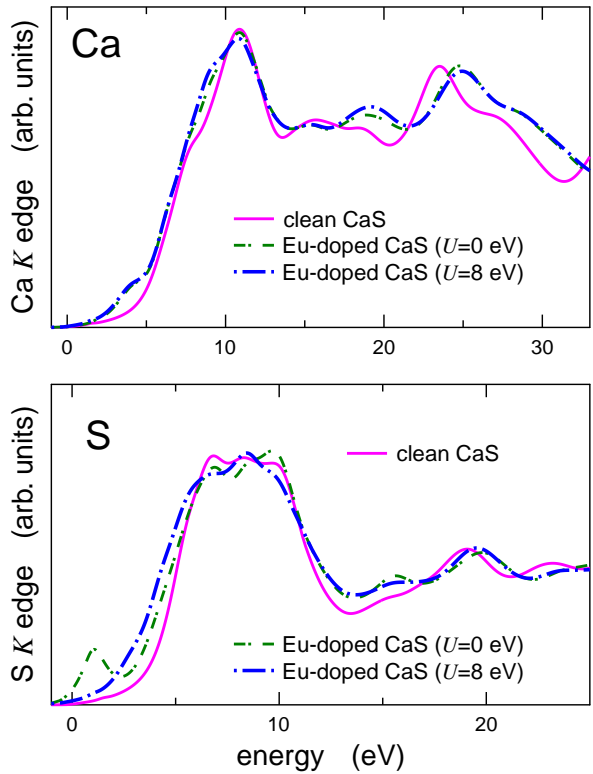}
  \label{fig-eucas}
\end{figure}

A lot of interest has been focused on properties of CaS (and other
sulfides) doped with rare earth metals. Fig.~\ref{fig-eucas} shows how
Ca and S \ke\ spectra of CaS change upon doping with Eu.  We present
the spectra generated at those atoms of given type which are closest
to the Eu impurity because here the effect will be largest.  It can be
seen that the presence of Eu atoms does not change the spectra very
much.  Apart from the region at the very edge, no new features are
introduced; just the intensities and positions of existing peaks and
shoulders change slightly.

Some pre-edge structure appears for both edges upon Eu doping; for the
Ca \ke\ this pre-peak occurs both for the LDA and LDA+$U$ calculations
while for the S \ke\ it occurs only for the LDA case.  It results from
hybridization of Ca and S states with Eu states.  The position of Eu
states depends on $U$, it is thus not surprising that the appearance
of the S \ke\ pre-peak is strongly $U$-dependent.

Experimental XANES of doped CaS would be a superposition of spectra
generated at all Ca or S atoms in the sample.  Only a small fraction
of them would occur near any Eu impurity.  Given the overall
similarity of the spectra for clean and doped CaS
(Fig.~\ref{fig-eucas}), it is likely that the Eu doping will not
affect the Ca and S \ke\ XANES of CaS significantly --- apart from a
pre-edge structure which is not present for clean CaS but which might
occur for doped CaS.


\section{Discussion}

Our goal was to check the robustness of XANES calculations on a case
study of Ca and S \ke\ spectra of CaS and to disentangle the effects
of different physical approximations from the effects of using
different computational techniques.  We found that Green's function
multiple-scattering calculations done by the \kkr\ code, FLAPW
calculations done by the \wtk\ code, and FDM calculations done by the
{\sc fdmnes} code yield similar results.  The experimental spectra
were reproduced quite well but there are some differences within the
first 10~eV above the edge.  Employing the full potential does not
lead to significantly different spectra in comparison with the ASA.
Including the core hole by the final state approximation enhances some
features close to the edge but generally does not lead to big changes
of the spectra with respect to the ground state calculations.

Earlier calculations of Xu \ea\ \cite{XLC+13} did not reproduce small
but well-defined peaks which appear in the experimental spectra at
35~eV above the Ca edge and at 23~eV above the S edge.  The
calculations of Kravtsova \ea\ \cite{KSS+04} did not reproduce the peak at
23~eV above the S \ke\ either.  Xu \ea\ \cite{XLC+13} tentatively
attributed these peaks to radiation damage or to an unspecified charge
transfer effect.  However, all our calculations reproduce these peaks
properly (cf.\ Fig.~\ref{fig-exper}).  It seems, therefore, that the
failure of earlier calculations to obtain these peaks was due to some
technical reasons.  One of them might be that perhaps a too large
broadening related to the finite photoelectron lifetime was applied.
One should bear in mind that the values for the electron mean free
path published by M\"{u}ller \ea \cite{MJW82} which are still commonly used
were obtained by summarizing the trends for a large set of data for
various materials; a large scatter of individual values around these
estimates thus has to be anticipated \cite{Gud+74,LS+74}.  Setting the
photoelectron lifetime broadening empirically {\em ad~hoc} probably
still remains the most practical way to deal with this issue, despite
promising results in the field of ab-initio mean free path calculation
\cite{KSP+07,CB+14,EKG+17}.

There is also a difference between our results and the results of Xu
\ea~\cite{XLC+13} concerning the importance of using the full
potential.  Our Green's function multiple-scattering calculations
indicate that using the full potential does not significantly change
the spectra with respect to the ASA results.  The FMD calculations of
Xu \ea\ \cite{XLC+13}, on the other hand, suggest that using the full
potential significantly improves the results within the first 10~eV
above the edge --- at least for the Ca \ke\ spectrum.  A possible
explanation of this apparent contradiction could be that the spherical
potential calculations done via the \kkr\ code relied on the ASA while
the spherical potential calculations of Xu \ea\ \cite{XLC+13} relied on
the muffin-tin approximation.  It has been argued that the ASA is
better than the pure muffin-tin approximation~\cite{ZA+09}.


\section{Conclusions}

Theoretical Ca and S \ke\ XANES spectra of CaS obtained by different
full-potential methods (Green's function multiple-scattering, FLAPW,
FDM) are mostly quite similar and in a good agreement with experiment.
In particular all calculations reproduce the small but well-defined
peaks which appear in the experimental spectra at 35~eV above the Ca
edge and at 23~eV above the S edge and which were not described by
earlier calculations.  Some differences between theory and experiment
remain within the first 10~eV above the edge, where the
finite-difference method ({\sc fdmnes} code) leads to better agreement
with experiment than the Green's function multiple-scattering method
({\sc sprkkr} code) or the FLAPW method ({\sc wien2k} code).

Full-potential and atomic sphere approximation (ASA) calculations done
within the Green's function multiple-scattering framework lead to very
similar spectra.  Including the core hole via the final state
approximation does not change the spectra significantly but it leads
to enhancement of features close to the absorption edge.  The
influence of the core hole on the electronic structure of atoms next
to the photoabsorbing atom can be neglected.  Doping CaS with Eu will
not affect the Ca and S \ke\ XANES of CaS significantly.


\begin{acknowledgements}
This work was supported by the GA~\v{C}R project 17-14840~S and
  by the CEDAMNF project CZ.02.1.01/0.0/0.0/15\_003/0000358.
\end{acknowledgements}



\begin{thebibliography}{35}%
\makeatletter
\providecommand \@ifxundefined [1]{%
 \@ifx{#1\undefined}
}%
\providecommand \@ifnum [1]{%
 \ifnum #1\expandafter \@firstoftwo
 \else \expandafter \@secondoftwo
 \fi
}%
\providecommand \@ifx [1]{%
 \ifx #1\expandafter \@firstoftwo
 \else \expandafter \@secondoftwo
 \fi
}%
\providecommand \natexlab [1]{#1}%
\providecommand \enquote  [1]{``#1''}%
\providecommand \bibnamefont  [1]{#1}%
\providecommand \bibfnamefont [1]{#1}%
\providecommand \citenamefont [1]{#1}%
\providecommand \href@noop [0]{\@secondoftwo}%
\providecommand \href [0]{\begingroup \@sanitize@url \@href}%
\providecommand \@href[1]{\@@startlink{#1}\@@href}%
\providecommand \@@href[1]{\endgroup#1\@@endlink}%
\providecommand \@sanitize@url [0]{\catcode `\\12\catcode `\$12\catcode
  `\&12\catcode `\#12\catcode `\^12\catcode `\_12\catcode `\%12\relax}%
\providecommand \@@startlink[1]{}%
\providecommand \@@endlink[0]{}%
\providecommand \url  [0]{\begingroup\@sanitize@url \@url }%
\providecommand \@url [1]{\endgroup\@href {#1}{\urlprefix }}%
\providecommand \urlprefix  [0]{URL }%
\providecommand \Eprint [0]{\href }%
\providecommand \doibase [0]{http://dx.doi.org/}%
\providecommand \selectlanguage [0]{\@gobble}%
\providecommand \bibinfo  [0]{\@secondoftwo}%
\providecommand \bibfield  [0]{\@secondoftwo}%
\providecommand \translation [1]{[#1]}%
\providecommand \BibitemOpen [0]{}%
\providecommand \bibitemStop [0]{}%
\providecommand \bibitemNoStop [0]{.\EOS\space}%
\providecommand \EOS [0]{\spacefactor3000\relax}%
\providecommand \BibitemShut  [1]{\csname bibitem#1\endcsname}%
\let\auto@bib@innerbib\@empty
\bibitem [{\citenamefont {Tamura}\ and\ \citenamefont
  {Shibukawa}(1993)}]{TS+93}%
  \BibitemOpen
  \bibfield  {author} {\bibinfo {author} {\bibfnamefont {Y.}~\bibnamefont
  {Tamura}}\ and\ \bibinfo {author} {\bibfnamefont {A.}~\bibnamefont
  {Shibukawa}},\ }\href@noop {} {\bibfield  {journal} {\bibinfo  {journal}
  {Jpn. J. Appl. Phys.}\ }\textbf {\bibinfo {volume} {32}},\ \bibinfo {pages}
  {3187} (\bibinfo {year} {1993})}\BibitemShut {NoStop}%
\bibitem [{\citenamefont {Zhang}\ \emph {et~al.}(2005)\citenamefont {Zhang},
  \citenamefont {Liang}, \citenamefont {Zhang},\ and\ \citenamefont
  {Su}}]{ZLZ+05}%
  \BibitemOpen
  \bibfield  {author} {\bibinfo {author} {\bibfnamefont {X.}~\bibnamefont
  {Zhang}}, \bibinfo {author} {\bibfnamefont {L.}~\bibnamefont {Liang}},
  \bibinfo {author} {\bibfnamefont {J.}~\bibnamefont {Zhang}}, \ and\ \bibinfo
  {author} {\bibfnamefont {Q.}~\bibnamefont {Su}},\ }\href@noop {} {\bibfield
  {journal} {\bibinfo  {journal} {Mater. Lett.}\ }\textbf {\bibinfo {volume}
  {59}},\ \bibinfo {pages} {749} (\bibinfo {year} {2005})}\BibitemShut
  {NoStop}%
\bibitem [{\citenamefont {Jia}\ and\ \citenamefont {Wang}(2007)}]{JW+07}%
  \BibitemOpen
  \bibfield  {author} {\bibinfo {author} {\bibfnamefont {D.}~\bibnamefont
  {Jia}}\ and\ \bibinfo {author} {\bibfnamefont {X.}~\bibnamefont {Wang}},\
  }\href@noop {} {\bibfield  {journal} {\bibinfo  {journal} {Opt. Mater.}\
  }\textbf {\bibinfo {volume} {30}},\ \bibinfo {pages} {375} (\bibinfo {year}
  {2007})}\BibitemShut {NoStop}%
\bibitem [{\citenamefont {Huang}(2015)}]{Hua+15}%
  \BibitemOpen
  \bibfield  {author} {\bibinfo {author} {\bibfnamefont {B.}~\bibnamefont
  {Huang}},\ }\href@noop {} {\bibfield  {journal} {\bibinfo  {journal} {Inorg.
  Chem.}\ }\textbf {\bibinfo {volume} {54}},\ \bibinfo {pages} {11423}
  (\bibinfo {year} {2015})}\BibitemShut {NoStop}%
\bibitem [{\citenamefont {Farrell}\ \emph {et~al.}(2002)\citenamefont
  {Farrell}, \citenamefont {Fleet}, \citenamefont {Stekhin}, \citenamefont
  {Kravtsova}, \citenamefont {Soldatov},\ and\ \citenamefont {Liu}}]{FFS+02}%
  \BibitemOpen
  \bibfield  {author} {\bibinfo {author} {\bibfnamefont {S.~P.}\ \bibnamefont
  {Farrell}}, \bibinfo {author} {\bibfnamefont {M.~E.}\ \bibnamefont {Fleet}},
  \bibinfo {author} {\bibfnamefont {I.~E.}\ \bibnamefont {Stekhin}}, \bibinfo
  {author} {\bibfnamefont {A.}~\bibnamefont {Kravtsova}}, \bibinfo {author}
  {\bibfnamefont {A.~V.}\ \bibnamefont {Soldatov}}, \ and\ \bibinfo {author}
  {\bibfnamefont {X.}~\bibnamefont {Liu}},\ }\href@noop {} {\bibfield
  {journal} {\bibinfo  {journal} {Am. Mineral.}\ }\textbf {\bibinfo {volume}
  {87}},\ \bibinfo {pages} {1321} (\bibinfo {year} {2002})}\BibitemShut
  {NoStop}%
\bibitem [{\citenamefont {Kravtsova}\ \emph {et~al.}(2004)\citenamefont
  {Kravtsova}, \citenamefont {Stekhin}, \citenamefont {Soldatov}, \citenamefont
  {Liu},\ and\ \citenamefont {Fleet}}]{KSS+04}%
  \BibitemOpen
  \bibfield  {author} {\bibinfo {author} {\bibfnamefont {A.~N.}\ \bibnamefont
  {Kravtsova}}, \bibinfo {author} {\bibfnamefont {I.~E.}\ \bibnamefont
  {Stekhin}}, \bibinfo {author} {\bibfnamefont {A.~V.}\ \bibnamefont
  {Soldatov}}, \bibinfo {author} {\bibfnamefont {X.}~\bibnamefont {Liu}}, \
  and\ \bibinfo {author} {\bibfnamefont {M.~E.}\ \bibnamefont {Fleet}},\
  }\href@noop {} {\bibfield  {journal} {\bibinfo  {journal} {Phys. Rev. B}\
  }\textbf {\bibinfo {volume} {69}},\ \bibinfo {pages} {134109} (\bibinfo
  {year} {2004})}\BibitemShut {NoStop}%
\bibitem [{\citenamefont {{Alonso Mori}}\ \emph {et~al.}(2009)\citenamefont
  {{Alonso Mori}}, \citenamefont {Paris}, \citenamefont {Giuli}, \citenamefont
  {Eeckhout}, \citenamefont {Kav\v{c}i\v{c}}, \citenamefont {\v{Z}itnik},
  \citenamefont {Bu\v{c}ar}, \citenamefont {Pettersson},\ and\ \citenamefont
  {Glatzel}}]{APG+09}%
  \BibitemOpen
  \bibfield  {author} {\bibinfo {author} {\bibfnamefont {R.}~\bibnamefont
  {{Alonso Mori}}}, \bibinfo {author} {\bibfnamefont {E.}~\bibnamefont
  {Paris}}, \bibinfo {author} {\bibfnamefont {G.}~\bibnamefont {Giuli}},
  \bibinfo {author} {\bibfnamefont {S.~G.}\ \bibnamefont {Eeckhout}}, \bibinfo
  {author} {\bibfnamefont {M.}~\bibnamefont {Kav\v{c}i\v{c}}}, \bibinfo
  {author} {\bibfnamefont {M.}~\bibnamefont {\v{Z}itnik}}, \bibinfo {author}
  {\bibfnamefont {K.}~\bibnamefont {Bu\v{c}ar}}, \bibinfo {author}
  {\bibfnamefont {L.~G.~M.}\ \bibnamefont {Pettersson}}, \ and\ \bibinfo
  {author} {\bibfnamefont {P.}~\bibnamefont {Glatzel}},\ }\href@noop {}
  {\bibfield  {journal} {\bibinfo  {journal} {Anal. Chem.}\ }\textbf {\bibinfo
  {volume} {81}},\ \bibinfo {pages} {6516} (\bibinfo {year}
  {2009})}\BibitemShut {NoStop}%
\bibitem [{\citenamefont {Xu}\ \emph {et~al.}(2013)\citenamefont {Xu},
  \citenamefont {Liu}, \citenamefont {Cui}, \citenamefont {Zheng},
  \citenamefont {Hu}, \citenamefont {Marcelli},\ and\ \citenamefont
  {Wu}}]{XLC+13}%
  \BibitemOpen
  \bibfield  {author} {\bibinfo {author} {\bibfnamefont {W.}~\bibnamefont
  {Xu}}, \bibinfo {author} {\bibfnamefont {L.}~\bibnamefont {Liu}}, \bibinfo
  {author} {\bibfnamefont {M.}~\bibnamefont {Cui}}, \bibinfo {author}
  {\bibfnamefont {L.}~\bibnamefont {Zheng}}, \bibinfo {author} {\bibfnamefont
  {Y.}~\bibnamefont {Hu}}, \bibinfo {author} {\bibfnamefont {A.}~\bibnamefont
  {Marcelli}}, \ and\ \bibinfo {author} {\bibfnamefont {Z.}~\bibnamefont
  {Wu}},\ }\href {\doibase 10.1107/S0909049512040617} {\bibfield  {journal}
  {\bibinfo  {journal} {Journal of Synchrotron Radiation}\ }\textbf {\bibinfo
  {volume} {20}},\ \bibinfo {pages} {110} (\bibinfo {year} {2013})}\BibitemShut
  {NoStop}%
\bibitem [{\citenamefont {Joly}(2015)}]{fdmnes-code}%
  \BibitemOpen
  \bibfield  {author} {\bibinfo {author} {\bibfnamefont {Y.}~\bibnamefont
  {Joly}},\ }\href@noop {} {\emph {\bibinfo {title} {The {\sc fdmnes} code}}},\
  \bibinfo {address} {\url{http://neel.cnrs.fr/spip.php?rubrique1007&lang=en}}
  (\bibinfo {year} {2015})\BibitemShut {NoStop}%
\bibitem [{\citenamefont {Bun\u{a}u}\ and\ \citenamefont {Joly}(2009)}]{BJ+09}%
  \BibitemOpen
  \bibfield  {author} {\bibinfo {author} {\bibfnamefont {O.}~\bibnamefont
  {Bun\u{a}u}}\ and\ \bibinfo {author} {\bibfnamefont {Y.}~\bibnamefont
  {Joly}},\ }\href {http://stacks.iop.org/0953-8984/21/i=34/a=345501}
  {\bibfield  {journal} {\bibinfo  {journal} {J. Phys.: Condens. Matter}\
  }\textbf {\bibinfo {volume} {21}},\ \bibinfo {pages} {345501} (\bibinfo
  {year} {2009})}\BibitemShut {NoStop}%
\bibitem [{\citenamefont {Rehr}(2013)}]{feff-code}%
  \BibitemOpen
  \bibfield  {author} {\bibinfo {author} {\bibfnamefont {J.~J.}\ \bibnamefont
  {Rehr}},\ }\href@noop {} {\emph {\bibinfo {title} {The {\sc feff} code,
  version 9}}},\ \bibinfo {address} {\url{http://feffproject.org}} (\bibinfo
  {year} {2013})\BibitemShut {NoStop}%
\bibitem [{\citenamefont {Rehr}\ \emph {et~al.}(2010)\citenamefont {Rehr},
  \citenamefont {Kas}, \citenamefont {Vila}, \citenamefont {Prange},\ and\
  \citenamefont {Jorissen}}]{RKV+10}%
  \BibitemOpen
  \bibfield  {author} {\bibinfo {author} {\bibfnamefont {J.~J.}\ \bibnamefont
  {Rehr}}, \bibinfo {author} {\bibfnamefont {J.~J.}\ \bibnamefont {Kas}},
  \bibinfo {author} {\bibfnamefont {F.~D.}\ \bibnamefont {Vila}}, \bibinfo
  {author} {\bibfnamefont {M.~P.}\ \bibnamefont {Prange}}, \ and\ \bibinfo
  {author} {\bibfnamefont {K.}~\bibnamefont {Jorissen}},\ }\href@noop {}
  {\bibfield  {journal} {\bibinfo  {journal} {Phys. Chem. Chem. Phys.}\
  }\textbf {\bibinfo {volume} {12}},\ \bibinfo {pages} {5503} (\bibinfo {year}
  {2010})}\BibitemShut {NoStop}%
\bibitem [{\citenamefont {Huhne}\ and\ \citenamefont {Ebert}(1999)}]{HE+99}%
  \BibitemOpen
  \bibfield  {author} {\bibinfo {author} {\bibfnamefont {T.}~\bibnamefont
  {Huhne}}\ and\ \bibinfo {author} {\bibfnamefont {H.}~\bibnamefont {Ebert}},\
  }\href {\doibase https://doi.org/10.1016/S0038-1098(98)00628-0} {\bibfield
  {journal} {\bibinfo  {journal} {Solid State Communications}\ }\textbf
  {\bibinfo {volume} {109}},\ \bibinfo {pages} {577} (\bibinfo {year}
  {1999})}\BibitemShut {NoStop}%
\bibitem [{\citenamefont {Dorenbos}(2005)}]{Dor+05}%
  \BibitemOpen
  \bibfield  {author} {\bibinfo {author} {\bibfnamefont {P.}~\bibnamefont
  {Dorenbos}},\ }\href@noop {} {\bibfield  {journal} {\bibinfo  {journal} {J.
  Phys.: Condens. Matter}\ }\textbf {\bibinfo {volume} {17}},\ \bibinfo {pages}
  {8103} (\bibinfo {year} {2005})}\BibitemShut {NoStop}%
\bibitem [{\citenamefont {Ebert}(2017)}]{sprkkr-code}%
  \BibitemOpen
  \bibfield  {author} {\bibinfo {author} {\bibfnamefont {H.}~\bibnamefont
  {Ebert}},\ }\href@noop {} {\emph {\bibinfo {title} {The {\sc sprkkr} code,
  version 7.7}}},\ \bibinfo {address}
  {\url{http://ebert.cup.uni-muenchen.de/SPRKKR}} (\bibinfo {year}
  {2017})\BibitemShut {NoStop}%
\bibitem [{\citenamefont {Ebert}\ \emph {et~al.}(2011)\citenamefont {Ebert},
  \citenamefont {K\"{o}dderitzsch},\ and\ \citenamefont {Min\'{a}r}}]{EKM11}%
  \BibitemOpen
  \bibfield  {author} {\bibinfo {author} {\bibfnamefont {H.}~\bibnamefont
  {Ebert}}, \bibinfo {author} {\bibfnamefont {D.}~\bibnamefont
  {K\"{o}dderitzsch}}, \ and\ \bibinfo {author} {\bibfnamefont
  {J.}~\bibnamefont {Min\'{a}r}},\ }\href@noop {} {\bibfield  {journal}
  {\bibinfo  {journal} {Rep. Prog. Phys.}\ }\textbf {\bibinfo {volume} {74}},\
  \bibinfo {pages} {096501} (\bibinfo {year} {2011})}\BibitemShut {NoStop}%
\bibitem [{\citenamefont {Blaha}\ \emph {et~al.}(2001)\citenamefont {Blaha},
  \citenamefont {Schwarz}, \citenamefont {Madsen}, \citenamefont {Kvasnicka},\
  and\ \citenamefont {Luitz}}]{Blaha+01}%
  \BibitemOpen
  \bibfield  {author} {\bibinfo {author} {\bibfnamefont {P.}~\bibnamefont
  {Blaha}}, \bibinfo {author} {\bibfnamefont {K.}~\bibnamefont {Schwarz}},
  \bibinfo {author} {\bibfnamefont {G.~K.~H.}\ \bibnamefont {Madsen}}, \bibinfo
  {author} {\bibfnamefont {D.}~\bibnamefont {Kvasnicka}}, \ and\ \bibinfo
  {author} {\bibfnamefont {J.}~\bibnamefont {Luitz}},\ }\href@noop {} {\emph
  {\bibinfo {title} {Wien2k, An Augmented Plane Wave plus Local orbital Program
  for Calculating the Crystal Properties}}},\ \bibinfo {address}
  {\url{http://www.wien2k.at}} (\bibinfo {year} {2001})\BibitemShut {NoStop}%
\bibitem [{\citenamefont {Zwierzycki}\ and\ \citenamefont
  {Andersen}(2009)}]{ZA+09}%
  \BibitemOpen
  \bibfield  {author} {\bibinfo {author} {\bibfnamefont {M.}~\bibnamefont
  {Zwierzycki}}\ and\ \bibinfo {author} {\bibfnamefont {O.}~\bibnamefont
  {Andersen}},\ }\href@noop {} {\bibfield  {journal} {\bibinfo  {journal} {Acta
  Phys. Pol. A}\ }\textbf {\bibinfo {volume} {115}},\ \bibinfo {pages} {64}
  (\bibinfo {year} {2009})}\BibitemShut {NoStop}%
\bibitem [{\citenamefont {Zeller}(1987)}]{Zel87a}%
  \BibitemOpen
  \bibfield  {author} {\bibinfo {author} {\bibfnamefont {R.}~\bibnamefont
  {Zeller}},\ }\href {http://dx.doi.org/10.1088/0022-3719/20/16/010} {\bibfield
   {journal} {\bibinfo  {journal} {J. Phys. C: Solid State Phys.}\ }\textbf
  {\bibinfo {volume} {20}},\ \bibinfo {pages} {2347} (\bibinfo {year}
  {1987})}\BibitemShut {NoStop}%
\bibitem [{\citenamefont {Huhne}\ \emph {et~al.}(1998)\citenamefont {Huhne},
  \citenamefont {Zecha}, \citenamefont {Ebert}, \citenamefont {Dederichs},\
  and\ \citenamefont {Zeller}}]{HZE+98}%
  \BibitemOpen
  \bibfield  {author} {\bibinfo {author} {\bibfnamefont {T.}~\bibnamefont
  {Huhne}}, \bibinfo {author} {\bibfnamefont {C.}~\bibnamefont {Zecha}},
  \bibinfo {author} {\bibfnamefont {H.}~\bibnamefont {Ebert}}, \bibinfo
  {author} {\bibfnamefont {P.~H.}\ \bibnamefont {Dederichs}}, \ and\ \bibinfo
  {author} {\bibfnamefont {R.}~\bibnamefont {Zeller}},\ }\href {\doibase
  10.1103/PhysRevB.58.10236} {\bibfield  {journal} {\bibinfo  {journal} {Phys.
  Rev. B}\ }\textbf {\bibinfo {volume} {58}},\ \bibinfo {pages} {10236}
  (\bibinfo {year} {1998})}\BibitemShut {NoStop}%
\bibitem [{\citenamefont {Joly}(2001)}]{Jol+01}%
  \BibitemOpen
  \bibfield  {author} {\bibinfo {author} {\bibfnamefont {Y.}~\bibnamefont
  {Joly}},\ }\href@noop {} {\bibfield  {journal} {\bibinfo  {journal} {Phys.
  Rev. B}\ }\textbf {\bibinfo {volume} {63}},\ \bibinfo {pages} {125120}
  (\bibinfo {year} {2001})}\BibitemShut {NoStop}%
\bibitem [{\citenamefont {Hedin}\ and\ \citenamefont {Lundqvist}(1969)}]{HL69}%
  \BibitemOpen
  \bibfield  {author} {\bibinfo {author} {\bibfnamefont {L.}~\bibnamefont
  {Hedin}}\ and\ \bibinfo {author} {\bibfnamefont {S.}~\bibnamefont
  {Lundqvist}},\ }\href@noop {} {\bibfield  {journal} {\bibinfo  {journal}
  {Solid State Phys.}\ }\textbf {\bibinfo {volume} {23}},\ \bibinfo {pages} {1}
  (\bibinfo {year} {1969})}\BibitemShut {NoStop}%
\bibitem [{\citenamefont {Lee}\ and\ \citenamefont {Beni}(1977)}]{LB77}%
  \BibitemOpen
  \bibfield  {author} {\bibinfo {author} {\bibfnamefont {P.~A.}\ \bibnamefont
  {Lee}}\ and\ \bibinfo {author} {\bibfnamefont {G.}~\bibnamefont {Beni}},\
  }\href@noop {} {\bibfield  {journal} {\bibinfo  {journal} {Phys. Rev. B}\
  }\textbf {\bibinfo {volume} {15}},\ \bibinfo {pages} {2862} (\bibinfo {year}
  {1977})}\BibitemShut {NoStop}%
\bibitem [{\citenamefont {Campbell}\ and\ \citenamefont {Papp}(2001)}]{CP01}%
  \BibitemOpen
  \bibfield  {author} {\bibinfo {author} {\bibfnamefont {J.~L.}\ \bibnamefont
  {Campbell}}\ and\ \bibinfo {author} {\bibfnamefont {T.}~\bibnamefont
  {Papp}},\ }\href@noop {} {\bibfield  {journal} {\bibinfo  {journal} {At. Data
  Nucl. Data Tables}\ }\textbf {\bibinfo {volume} {7}},\ \bibinfo {pages} {1}
  (\bibinfo {year} {2001})}\BibitemShut {NoStop}%
\bibitem [{\citenamefont {M\"{u}ller}\ \emph {et~al.}(1982)\citenamefont
  {M\"{u}ller}, \citenamefont {Jepsen},\ and\ \citenamefont {Wilkins}}]{MJW82}%
  \BibitemOpen
  \bibfield  {author} {\bibinfo {author} {\bibfnamefont {J.~E.}\ \bibnamefont
  {M\"{u}ller}}, \bibinfo {author} {\bibfnamefont {O.}~\bibnamefont {Jepsen}},
  \ and\ \bibinfo {author} {\bibfnamefont {J.~W.}\ \bibnamefont {Wilkins}},\
  }\href@noop {} {\bibfield  {journal} {\bibinfo  {journal} {Solid State
  Commun.}\ }\textbf {\bibinfo {volume} {42}},\ \bibinfo {pages} {365}
  (\bibinfo {year} {1982})}\BibitemShut {NoStop}%
\bibitem [{\citenamefont {Min\'ar}\ \emph {et~al.}(2006)\citenamefont
  {Min\'ar}, \citenamefont {Bornemann}, \citenamefont {\v{S}ipr}, \citenamefont
  {Polesya},\ and\ \citenamefont {Ebert}}]{MBS+06}%
  \BibitemOpen
  \bibfield  {author} {\bibinfo {author} {\bibfnamefont {J.}~\bibnamefont
  {Min\'ar}}, \bibinfo {author} {\bibfnamefont {S.}~\bibnamefont {Bornemann}},
  \bibinfo {author} {\bibfnamefont {O.}~\bibnamefont {\v{S}ipr}}, \bibinfo
  {author} {\bibfnamefont {S.}~\bibnamefont {Polesya}}, \ and\ \bibinfo
  {author} {\bibfnamefont {H.}~\bibnamefont {Ebert}},\ }\href@noop {}
  {\bibfield  {journal} {\bibinfo  {journal} {Appl. Physics A}\ }\textbf
  {\bibinfo {volume} {82}},\ \bibinfo {pages} {139} (\bibinfo {year}
  {2006})}\BibitemShut {NoStop}%
\bibitem [{\citenamefont {Zeller}(1988)}]{Zel88}%
  \BibitemOpen
  \bibfield  {author} {\bibinfo {author} {\bibfnamefont {R.}~\bibnamefont
  {Zeller}},\ }\href {\doibase 10.1007/BF01313114} {\bibfield  {journal}
  {\bibinfo  {journal} {Z. Physik B}\ }\textbf {\bibinfo {volume} {72}},\
  \bibinfo {pages} {79} (\bibinfo {year} {1988})}\BibitemShut {NoStop}%
\bibitem [{\citenamefont {Weijs}\ \emph {et~al.}(1990)\citenamefont {Weijs},
  \citenamefont {Czy\ifmmode~\dot{z}\else \.{z}\fi{}yk}, \citenamefont {van
  Acker}, \citenamefont {Speier}, \citenamefont {Goedkoop}, \citenamefont {van
  Leuken}, \citenamefont {Hendrix}, \citenamefont {de~Groot}, \citenamefont
  {van~der Laan}, \citenamefont {Buschow}, \citenamefont {Wiech},\ and\
  \citenamefont {Fuggle}}]{WCA+90}%
  \BibitemOpen
  \bibfield  {author} {\bibinfo {author} {\bibfnamefont {P.~J.~W.}\
  \bibnamefont {Weijs}}, \bibinfo {author} {\bibfnamefont {M.~T.}\ \bibnamefont
  {Czy\ifmmode~\dot{z}\else \.{z}\fi{}yk}}, \bibinfo {author} {\bibfnamefont
  {J.~F.}\ \bibnamefont {van Acker}}, \bibinfo {author} {\bibfnamefont
  {W.}~\bibnamefont {Speier}}, \bibinfo {author} {\bibfnamefont {J.~B.}\
  \bibnamefont {Goedkoop}}, \bibinfo {author} {\bibfnamefont {H.}~\bibnamefont
  {van Leuken}}, \bibinfo {author} {\bibfnamefont {H.~J.~M.}\ \bibnamefont
  {Hendrix}}, \bibinfo {author} {\bibfnamefont {R.~A.}\ \bibnamefont
  {de~Groot}}, \bibinfo {author} {\bibfnamefont {G.}~\bibnamefont {van~der
  Laan}}, \bibinfo {author} {\bibfnamefont {K.~H.~J.}\ \bibnamefont {Buschow}},
  \bibinfo {author} {\bibfnamefont {G.}~\bibnamefont {Wiech}}, \ and\ \bibinfo
  {author} {\bibfnamefont {J.~C.}\ \bibnamefont {Fuggle}},\ }\href {\doibase
  10.1103/PhysRevB.41.11899} {\bibfield  {journal} {\bibinfo  {journal} {Phys.
  Rev. B}\ }\textbf {\bibinfo {volume} {41}},\ \bibinfo {pages} {11899}
  (\bibinfo {year} {1990})}\BibitemShut {NoStop}%
\bibitem [{\citenamefont {\ifmmode~\check{S}\else \v{S}\fi{}ipr}\ \emph
  {et~al.}(1997)\citenamefont {\ifmmode~\check{S}\else \v{S}\fi{}ipr},
  \citenamefont {Machek}, \citenamefont {\ifmmode \check{S}\else
  \v{S}\fi{}im\ifmmode~\mathring{u}\else \r{u}\fi{}nek}, \citenamefont
  {Vack\'a\ifmmode~\check{r}\else \v{r}\fi{}},\ and\ \citenamefont
  {Hor\'ak}}]{SMS+97}%
  \BibitemOpen
  \bibfield  {author} {\bibinfo {author} {\bibfnamefont {O.}~\bibnamefont
  {\ifmmode~\check{S}\else \v{S}\fi{}ipr}}, \bibinfo {author} {\bibfnamefont
  {P.}~\bibnamefont {Machek}}, \bibinfo {author} {\bibfnamefont
  {A.}~\bibnamefont {\ifmmode \check{S}\else
  \v{S}\fi{}im\ifmmode~\mathring{u}\else \r{u}\fi{}nek}}, \bibinfo {author}
  {\bibfnamefont {J.}~\bibnamefont {Vack\'a\ifmmode~\check{r}\else
  \v{r}\fi{}}}, \ and\ \bibinfo {author} {\bibfnamefont {J.}~\bibnamefont
  {Hor\'ak}},\ }\href {\doibase 10.1103/PhysRevB.56.13151} {\bibfield
  {journal} {\bibinfo  {journal} {Phys. Rev. B}\ }\textbf {\bibinfo {volume}
  {56}},\ \bibinfo {pages} {13151} (\bibinfo {year} {1997})}\BibitemShut
  {NoStop}%
\bibitem [{\citenamefont {{\v{S}}ipr}\ and\ \citenamefont
  {Rocca}(2010)}]{SR+10}%
  \BibitemOpen
  \bibfield  {author} {\bibinfo {author} {\bibfnamefont {O.}~\bibnamefont
  {{\v{S}}ipr}}\ and\ \bibinfo {author} {\bibfnamefont {F.}~\bibnamefont
  {Rocca}},\ }\href {\doibase 10.1107/S0909049510008800} {\bibfield  {journal}
  {\bibinfo  {journal} {Journal of Synchrotron Radiation}\ }\textbf {\bibinfo
  {volume} {17}},\ \bibinfo {pages} {367} (\bibinfo {year} {2010})}\BibitemShut
  {NoStop}%
\bibitem [{\citenamefont {Gudat}(1974)}]{Gud+74}%
  \BibitemOpen
  \bibfield  {author} {\bibinfo {author} {\bibfnamefont {W.}~\bibnamefont
  {Gudat}},\ }\href@noop {} {Ph.D. thesis},\ \bibinfo  {school}
  {Universit\"{a}t Hamburg}, \bibinfo {address} {Hamburg} (\bibinfo {year}
  {1974})\BibitemShut {NoStop}%
\bibitem [{\citenamefont {Lindau}\ and\ \citenamefont {Spicer}(1974)}]{LS+74}%
  \BibitemOpen
  \bibfield  {author} {\bibinfo {author} {\bibfnamefont {I.}~\bibnamefont
  {Lindau}}\ and\ \bibinfo {author} {\bibfnamefont {W.}~\bibnamefont
  {Spicer}},\ }\href@noop {} {\bibfield  {journal} {\bibinfo  {journal}
  {Journal of Electron Spectroscopy and Related Phenomena}\ }\textbf {\bibinfo
  {volume} {3}},\ \bibinfo {pages} {409} (\bibinfo {year} {1974})}\BibitemShut
  {NoStop}%
\bibitem [{\citenamefont {Kas}\ \emph {et~al.}(2007)\citenamefont {Kas},
  \citenamefont {Sorini}, \citenamefont {Prange}, \citenamefont {Cambell},
  \citenamefont {Soininen},\ and\ \citenamefont {Rehr}}]{KSP+07}%
  \BibitemOpen
  \bibfield  {author} {\bibinfo {author} {\bibfnamefont {J.~J.}\ \bibnamefont
  {Kas}}, \bibinfo {author} {\bibfnamefont {A.~P.}\ \bibnamefont {Sorini}},
  \bibinfo {author} {\bibfnamefont {M.~P.}\ \bibnamefont {Prange}}, \bibinfo
  {author} {\bibfnamefont {L.~W.}\ \bibnamefont {Cambell}}, \bibinfo {author}
  {\bibfnamefont {J.~A.}\ \bibnamefont {Soininen}}, \ and\ \bibinfo {author}
  {\bibfnamefont {J.~J.}\ \bibnamefont {Rehr}},\ }\href@noop {} {\bibfield
  {journal} {\bibinfo  {journal} {Phys. Rev. B}\ }\textbf {\bibinfo {volume}
  {76}} (\bibinfo {year} {2007})}\BibitemShut {NoStop}%
\bibitem [{\citenamefont {Chantler}\ and\ \citenamefont
  {Bourke}(2014)}]{CB+14}%
  \BibitemOpen
  \bibfield  {author} {\bibinfo {author} {\bibfnamefont {C.~T.}\ \bibnamefont
  {Chantler}}\ and\ \bibinfo {author} {\bibfnamefont {J.~D.}\ \bibnamefont
  {Bourke}},\ }\href@noop {} {\bibfield  {journal} {\bibinfo  {journal} {J.
  Phys. Chem. A}\ }\textbf {\bibinfo {volume} {118}},\ \bibinfo {pages} {909}
  (\bibinfo {year} {2014})}\BibitemShut {NoStop}%
\bibitem [{\citenamefont {Emfietzoglou}\ \emph {et~al.}(2017)\citenamefont
  {Emfietzoglou}, \citenamefont {Kyriakou}, \citenamefont {Garcia-Molina},\
  and\ \citenamefont {Abril}}]{EKG+17}%
  \BibitemOpen
  \bibfield  {author} {\bibinfo {author} {\bibfnamefont {D.}~\bibnamefont
  {Emfietzoglou}}, \bibinfo {author} {\bibfnamefont {I.}~\bibnamefont
  {Kyriakou}}, \bibinfo {author} {\bibfnamefont {R.}~\bibnamefont
  {Garcia-Molina}}, \ and\ \bibinfo {author} {\bibfnamefont {I.}~\bibnamefont
  {Abril}},\ }\href@noop {} {\bibfield  {journal} {\bibinfo  {journal} {Surf.
  Interface Anal.}\ }\textbf {\bibinfo {volume} {49}},\ \bibinfo {pages} {4}
  (\bibinfo {year} {2017})}\BibitemShut {NoStop}%
\end{thebibliography}


%

\end{document}